# Comparison of the Magnetic properties of $Mn_3Fe_2Si_3O_{12}$ as a crystalline garnet and as a glass

F. Alex Cevallos and R.J. Cava

Department of Chemistry, Princeton University, Princeton NJ 08544

*Corresponding author: fac2@princeton.edu (F. Alex Cevallos)

**Abstract**

The crystalline garnet $Mn_3Fe_2Si_3O_{12}$ and an amorphous phase of the same nominal composition are synthesized at high pressure. The magnetic properties of the two forms are reported. Both phases order antiferromagnetically. The crystalline phase exhibits a Curie-Weiss theta of -47.2 K, with a sharp ordering transition at 12 K. The glassy phase exhibits a larger antiferromagnetic Curie-Weiss theta, of -83.0 K, with a broad ordering transition observed at 2.5 K. Both phases can be classified as magnetically frustrated, although the amorphous phase shows a much higher degree of frustration. The amorphous phase exhibits spin-glass behavior and is determined to have an actual composition of $Mn_3Fe_2Si_3O_{13}$.

**Introduction**

A geometrically frustrated magnet is a material whose structural geometry inhibits long-range magnetic ordering[1]. The archetypical examples are materials whose unpaired electrons interact antiferromagnetically and are on the corners of a triangular lattice, but other geometries can also lead to frustration of antiferromagnetically coupled spins. Random arrangements of spins in a solid can also lead to the inhibition of long range magnetic ordering due to the distribution of magnetic interactions present. In both cases, magnetic ordering often does not emerge until significantly below the Weiss temperature. For crystalline solids, one of the interactions often dominates and a long range ordered state is obtained at low temperatures, while in amorphous solids, a "spin glass" is often observed at low temperatures. Magnetically frustrated phases are of significant scientific interest, as long-range magnetic ordering is among the most common electronic ground states, and hindering it allows for the observation of a number of unusual and interesting properties[1-4]

The garnet crystal structure, which can be generalized as belonging to the form $A_3B_2C_3O_{12}$, has long served as an interesting materials platform. Garnets often easily form solid solutions, and a wide variety of atoms can fit on the *A*, *B*, and *C* sites, leading to an enormous number of possible combinations[5]. One particularly fruitful area of research has been the study of garnets containing magnetic atoms. These garnets can exhibit a wide variety of unusual properties, with examples of geometrically frustrated magnetism[6-7], ferrimagnetism[8], spin glass[6], spin ice[9], and spin liquid[2] reported. Geometric magnetic frustration can be found in garnets when the magnetic atoms decorate the A sublattice, as that lattice is made of corner sharing triangles[5].

While many garnets of interest occur in nature, many others are uncommon. In addition, many garnets form solid solutions of variable composition. One such example is the garnet calderite ($Mn_3Fe_2Si_3O_{12}$), which occurs naturally in solid solution with the garnet spessartine ($Mn_3Al_2Si_3O_{12}$); the pure Fe case has yet to be found in nature with no Al present[11]. As the formation conditions for calderite are of interest in minerology, there have been multiple publications on the synthesis and structure of this material[11-14], although a study of its magnetic properties has apparently not been previously reported. As the $Mn_3Fe_2Si_3O_{12}$ garnet contains two different magnetic atoms (i.e. $Mn^{2+}$ and $Fe^{3+}$) in distinct structural positions (i.e. the A and B sites in the garnet), we can expect potentially interesting magnetic properties due to a balance of magnetic interactions. Extending our previous work on the crystalline and amorphous forms of the spessartine garnet $Mn_2Al_3Si_2O_{12}$, which has one magnetic species present[15], we have been able obtain both crystalline and amorphous phases for the Mn-Fe-Si-O garnet system. We report here the basic magnetic properties of the calderite garnet, as well as the synthesis and basic magnetic properties of a previously unreported amorphous phase of the same nominal composition.

**Experimental**

*Synthesis*

Crystalline and amorphous forms of nominal composition $Mn_3Fe_2Si_3O_{12}$ were synthesized at a pressure of 6 GPa using a Rockland Research Corp. cubic multi-anvil cell. The amorphous phase was synthesized by grinding stoichiometric amounts of $Fe_2O_3$ (99.85%, Alfa Aesar), $SiO_2$ (99.8%, Alfa Aesar), and $MnO_2$ (99.8%, Fisher) with an agate mortar and pestle. The mixture was loaded into a boron nitride (BN) crucible and placed in a pyrophillite sample cube with a graphite furnace, which was pressurized to 6 GPa. The sample was then heated to 1100°C for 30 minutes, and quenched to room temperature. Small amounts of impurities were found to form at the sample-crucible boundary; these were mechanically removed and do not influence our measurements. The resulting sample was in the form of a hard, glossy pellet with a dark yellow-tinted brown color.

The crystalline garnet $Mn_3Fe_2Si_3O_{12}$ was synthesized according to the procedure reported by Lattard and Schreyer[11]. A stoichiometric mixture of $Fe_2O_3$ (99.85%, Alfa Aesar), $SiO_2$ (99.8%, Alfa Aesar), and $Mn_2O_3$ (99.99%, Cerac) was placed into a sealed gold crucible. The gold crucible was then placed within a larger BN crucible, and the space between them was packed with a mixture of $Fe_2O_3$ and $Fe_3O_4$ to act as an oxygen buffer. The crucible was placed in a sample cube and pressurized to 6 GPa. It was then heated to 700°C for 72 hours, before being quenched to room temperature. The resulting sample was an inhomogeneous mixture of the silicate garnet and a number of metal-oxide impurities. These impurities were removed by heating the sample in concentrated hydrochloric acid for 2 hours, and then washing with water. The resulting polycrystalline powder was a light yellow color, appearing almost white when ground.

*Characterization*

Room-temperature powder X-ray diffraction (PXRD) measurements were taken using a Bruker D8 Advance Eco diffractometer with a LynxEye-XE detector using Cu *Kα* radiation ($\lambda = 1.5418$ Å). Preliminary phase identification was performed using the Bruker EVA program. Zero-background measurements were taken on a polished Si wafer sample holder. A Rietveld refinement of the crystalline garnet was performed using the FullProf Suite. Magnetization measurements were taken with a Quantum Design Physical Property Measurement System (PPMS) Dynacool in the vibrating sample magnetometer configuration. The field-dependent magnetization for both forms was measured at 2 K. DC magnetic susceptibility measurements were taken between 1.8 K and 300 K in an applied field of 10,000 Oe, and the resulting data was fit using the Curie-Weiss Law. Thermogravimetric analysis was carried out using a TA Instruments SDT Q600, under a flowing atmosphere of 5% $H_2$ and 95% argon (Airgas).

**Results and Discussion**

The X-Ray diffraction pattern of the crystalline garnet $Mn_3Fe_2Si_3O_{12}$ was found to closely match previously reported patterns[13,16]. A small amount of impurity can be observed, which is believed to be coesite, a high-pressure polymorph of $SiO_2$. A Rietveld refinement on the sample was employed to determine a lattice constant *a* of 11.82239(2) Å, in good agreement with the published value of 11.8288 Å[13,16]. Other resulting structural parameters can be seen in Table 1. The XRD pattern of the glassy phase of nominal composition Mn$_3$Fe$_2$Si$_3$O$_{12}$ on a single crystal miscut silicon zero-background sample holder shows a large, broad hump, with no peaks corresponding to the crystalline garnet phase visible. One small, relatively broad peak can be observed near 27°, which appears to be a polymorph of SiO$_2$ and does not correspond to any known potential transition-metal-containing impurity phases. The results of the Rietveld refinement and the pattern of the glassy phase can be observed in Figure 1.

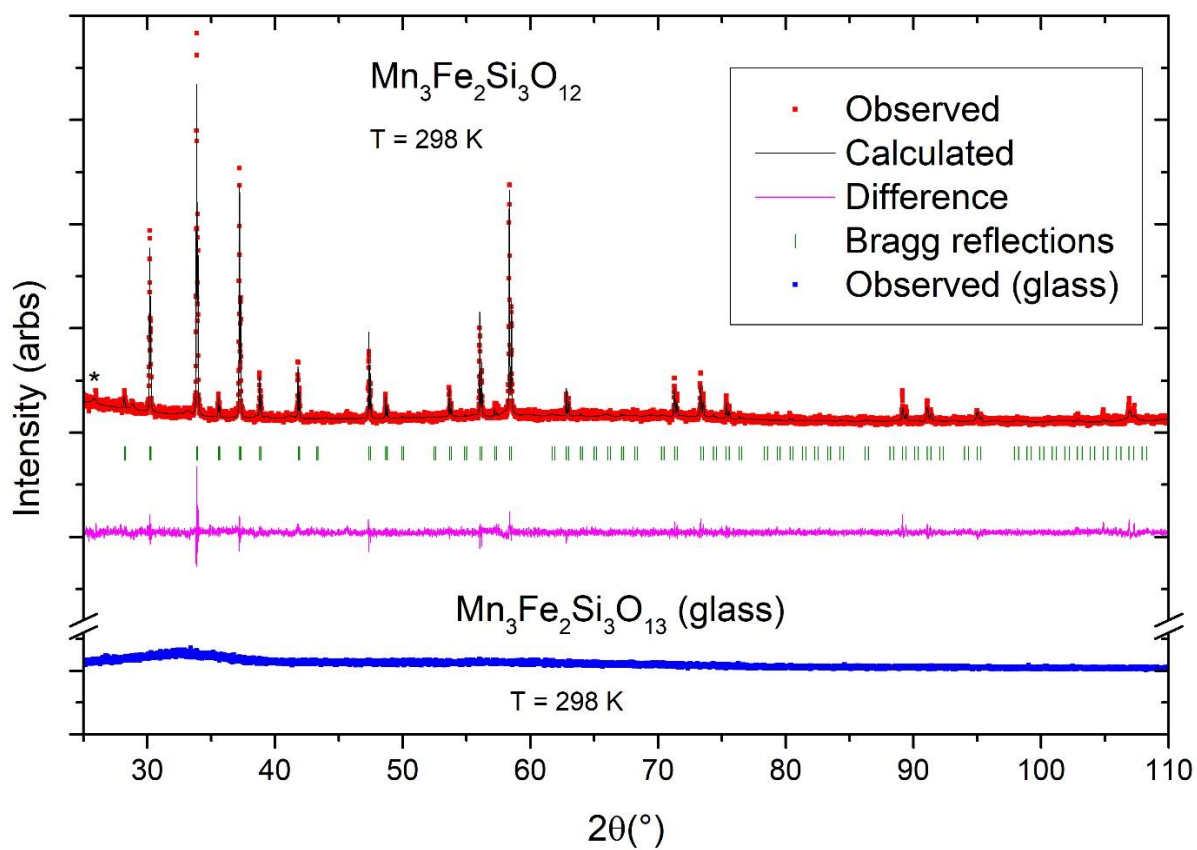

**Figure 1.** Rietveld refinement of the crystalline phase of Mn$_3$Fe$_2$Si$_3$O$_{12}$ (red) for 2θ between 25° and 110°. The pattern of the glassy phase is below, in blue. The small peak indicated with an asterisk corresponds to coesite (SiO$_2$). The region between 5° and 25° is obstructed by a large air scattering signal and has been removed from this image.

| Atom | Wyckoff site | x | y | z | $B_{iso}$ |
|---|---|---|---|---|---|
| $O^{2-}$ | 96h | 0.04249(50) | 0.05455(42) | 0.65636(75) | 2.71(15) |
| $Fe^{3+}$ | 16a | 0 | 0 | 0 | 3.63(13) |
| $Mn^{2+}$ | 24c | .125 | 0 | .25 | 3.16(9) |
| $Si^{4+}$ | 24d | .375 | 0 | .25 | 2.06(15) |

**Table 1**. Results of Rietveld refinement of the crystalline form of $Mn_3Fe_2Si_3O_{12}$ at ambient temperature, space group *Ia-3d* (No. 230), lattice parameter $a$ = 11.82239(2) Å. $R_f$ factor = 7.30, $R_p$ = 9.71, $R_{wp}$ = 12.4, $\chi^2$ = 1.28.

The temperature-dependent magnetic susceptibility and field dependent magnetization of the crystalline form of $Mn_3Fe_2Si_3O_{12}$ can be seen in Figure 2. The temperature-dependent susceptibility shows a clear magnetic transition, with a sharply-defined peak, at 12 K. The high-temperature region of the inverse susceptibility (200 K – 300 K) was fit to the Curie-Weiss law, $\chi - \chi_0 = C / (T - \theta_W)$, where $\chi$ is the magnetic susceptibility, $\chi_0$ is a temperature-independent contribution, C is the Curie Constant, and $\theta_W$ is the Weiss Temperature. Inverse susceptibility was found to be most linear for a temperature-independent contribution $\chi_0$ of -0.038 emu mol$^{-1}$. The resulting fit produces a Curie constant (C) of 20.7 and a Weiss temperature ($\theta_W$) of -49.0 K, suggesting that the observed peak in the susceptibility is an antiferromagnetic ordering transition. As there are 5 magnetic ions per formula unit, the effective magnetic moment per ion, $\mu_{eff}$, was determined by the relationship $\mu_{eff} = \sqrt{8\,C/5}$, and was thus found to be 5.8 $\mu_B$/ion, extremely close to the average of the spin-only values for the free ions ($Fe^{3+}$ and $Mn^{2+}$; 5.92 $\mu_B$/ion). It can be clearly observed that below the antiferromagnetic transition, as the temperature decreases, the susceptibility increases once more. This low temperature upturn is attributed to a very small number of paramagnetic impurities in the garnet. The susceptibility vs applied field at 2 K shows a slight curvature below 20,000 Oe but is essentially linear for all higher fields, with no signs of saturation.

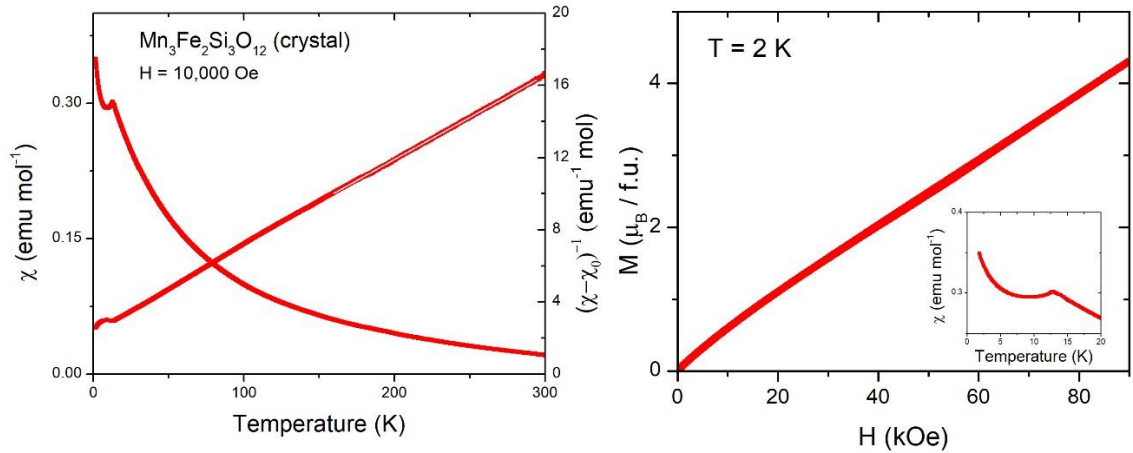

**Figure 2.** Left panel: The temperature-dependent DC magnetic susceptibility and reciprocal susceptibility of the crystalline garnet $Mn_3Fe_2Si_3O_{12}$ in an applied field of 10,000 Oe. A high-temperature Curie-Weiss fit is shown in white. Right panel: The field-dependent magnetization at 2 K. Inset: Magnified view of the magnetic susceptibility below 20 K.

The same measurements were performed on a sample of the amorphous phase. These are presented in Figure 3. Unlike the crystalline phase, which shows a sharp ordering transition at 12 K, the glassy phase has a broad transition, centered at approximately 2.5 K. A Curie-Weiss fit to the data in the region of 100 K – 300 K (with $\chi_0$ = -0.001 emu mol$^{-1}$) resulted in values of C = 13.0, $\theta_W$ = -82.9 K, and $\mu_{eff}$ = 4.6 $\mu_B$/ion. The Weiss temperature of the glass was significantly more negative than that of the crystal, a property which has previously been observed in the related system $Mn_3Al_2Si_3O_{12}$, and was attributed to superexchange-mediated interactions between $Mn^{2+}$ ions, and shorter Mn-Mn distances in the glass as compared to the crystal[15]. A look at the magnetic frustration index ($f = \theta_W / T_N$) of the two systems at 10,000 Oe shows that the glassy phase ($f$ = 33) is much more magnetically frustrated than the crystalline phase ($f$ = 4), with the crystalline phase only being somewhat frustrated in spite of its triangular network of Mn ions.

| | C | θ$_W$ | μ$_{eff}$ | T$_N$ |
|---|---|---|---|---|
| Crystalline | 20.7 | -49.0 K | 5.8 μ$_B$/ion | 12 K |
| Glassy | 13.0 | -82.9 K | 4.6 μ$_B$/ion | 2.5 K |

**Table 2**. Curie-Weiss values for temperature-dependent susceptibility of crystalline and glassy phases of nominal composition Mn$_3$Fe$_2$Si$_3$O$_{12}$ at 10000 Oe.

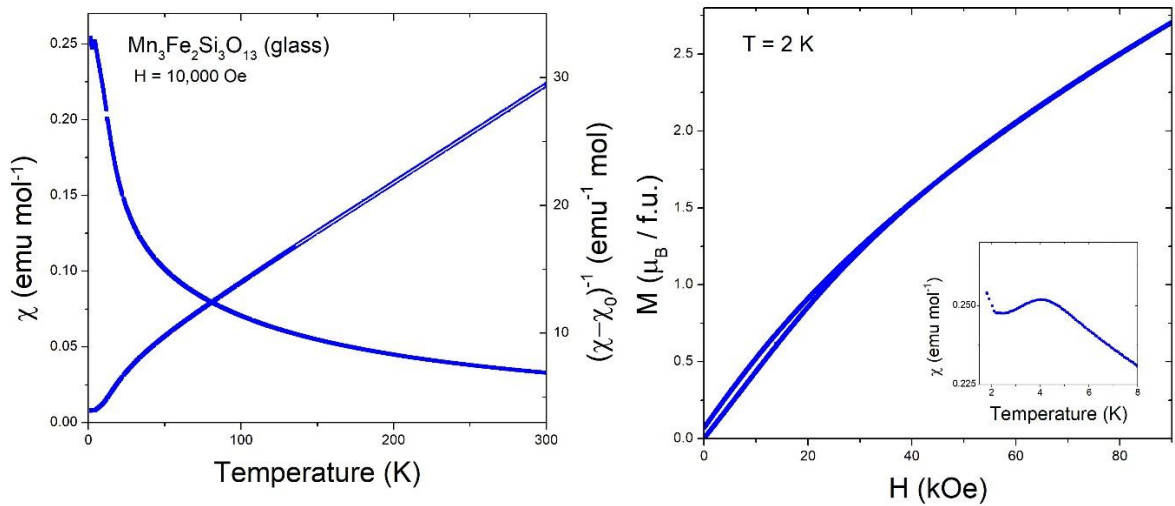

**Figure 3.** Left panel: The temperature-dependent DC magnetic susceptibility and reciprocal susceptibility of a glassy phase of composition Mn$_3$Fe$_2$Si$_3$O$_{13}$ in an applied field of 10,000 Oe. A high-temperature Curie-Weiss fit is shown in white. Right panel: The field-dependent magnetization at 2 K. Inset: Magnified view of the magnetic susceptibility below 8 K.

The significantly lower effective magnetic moment per ion of the glassy phase was an unexpected property of this material. As the structure is amorphous, it is possible that some of the metal atoms are not in the same oxidation state as in the garnet structure. To determine if this is the case, a thermogravimetric analysis was performed under a flowing atmosphere of 5% H$_2$/95% Ar, in the temperature range of 200-650°C. The results of this analysis can be seen in Figure 4. The resulting curve shows a gradual drop in the mass, followed by two sharper drops, for a total loss of approximately 2.8% mass. This mass loss corresponds almost exactly to the change between $Mn_3Fe_2Si_3O_{13}$ and $Mn_3Fe_2Si_3O_{12}$, suggesting that the glassy phase contains one additional oxygen atom per formula unit. Higher oxidation states of Manganese would lower the expected effective magnetic moment per ion; however, this change would not account for the difference observed. Assuming one $Mn^{4+}$ ion (consistent with the two clear drops in mass), the effective moment per ion would be expected to be 5.51 $\mu_B$ per ion, still significantly higher than the observed value of 4.6 $\mu_B$. The same value of 5.51 $\mu_B$ would be expected for the case of two $Mn^{3+}$ ions, suggesting that this discrepancy is likely not simply due to an increase in oxidation state in the glassy phase.

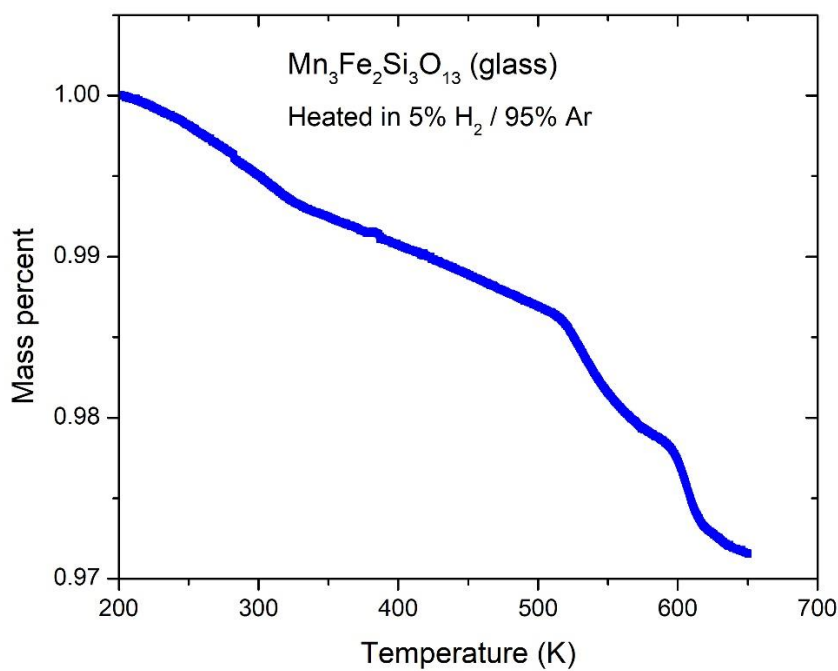

**Figure 4.** Thermogravimetric analysis of the glassy phase $Mn_3Fe_2Si_3O_{13}$, showing percentage of mass lost with heating under a flowing atmosphere of 5% $H_2$ in Argon.

In order to more directly compare the magnetic behavior of the two materials, the susceptibility graphs can be scaled by multiplying the inverse susceptibility by C / |θ_W|, and dividing T by θ_W. The

comparison of the unscaled and scaled inverse susceptibilities can be seen in Figure 5. As can be seen in the panel on the right, the temperature-dependent susceptibility of the glassy phase deflects away from ideal behavior at a higher relative temperature than that of the crystalline phase, above the glassy phase's magnetic transition. In contrast, the crystalline phase demonstrates nearly ideal Curie-Weiss behavior until the onset of the magnetic transition, which suggests that short range magnetic correlations, if present at all, are not substantial above the full three-dimensional magnetic ordering transition.

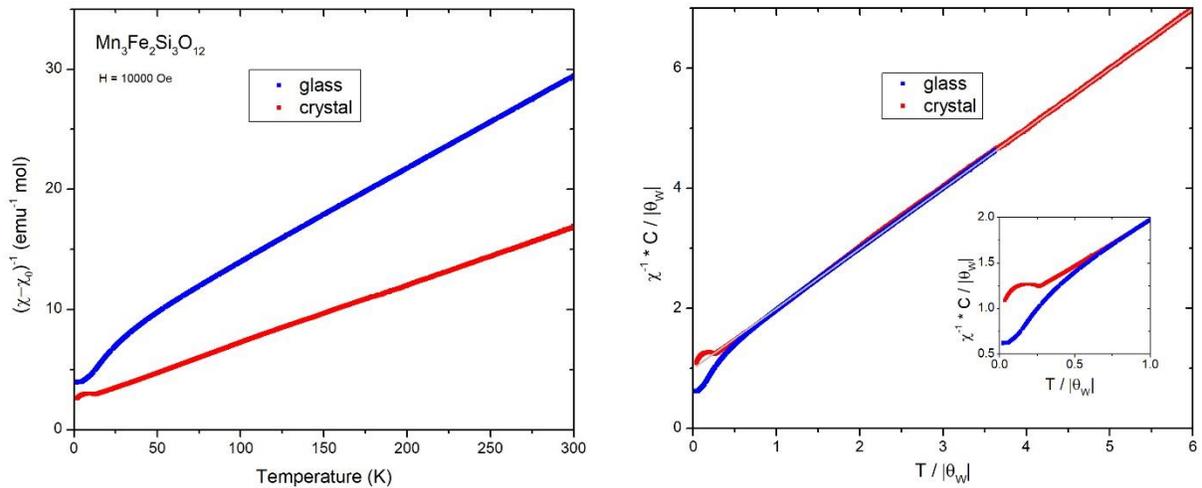

**Figure 5.** Left: Comparison of the inverse DC magnetic susceptibility of the crystalline phase $Mn_3Fe_2Si_3O_{12}$ (red) and the glassy phase $Mn_3Fe_2Si_3O_{13}$ (blue). Right: Scaled inverse susceptibilities, allowing for direct comparison of the crystalline and glassy phases. In grey is the ideal Curie-Weiss behavior, extrapolated from high-temperature data. Inset: The low-temperature region of the scaled inverse susceptibilities.

The zero-field-cooled and field-cooled DC magnetic susceptibility of the glassy phase were measured in an applied field of 1,000 Oe. The results, shown in Figure 6, show clear signs of spin-glass behavior. This behavior is indicative of a large degree of disorder in the magnetic spins, as is expected to arise from the amorphous structure of the glass. It is noted that the highest value of the susceptibility now occurs at approximately 7.5 K, instead of 2.5 K as in the measurement at 10,000 Oe. This is consisitent with expectations for dominantly antiferromagnetic spin-glass ordering. A similar measurement performed on the crystalline phase does not yield any indication of spin-glass behavior.

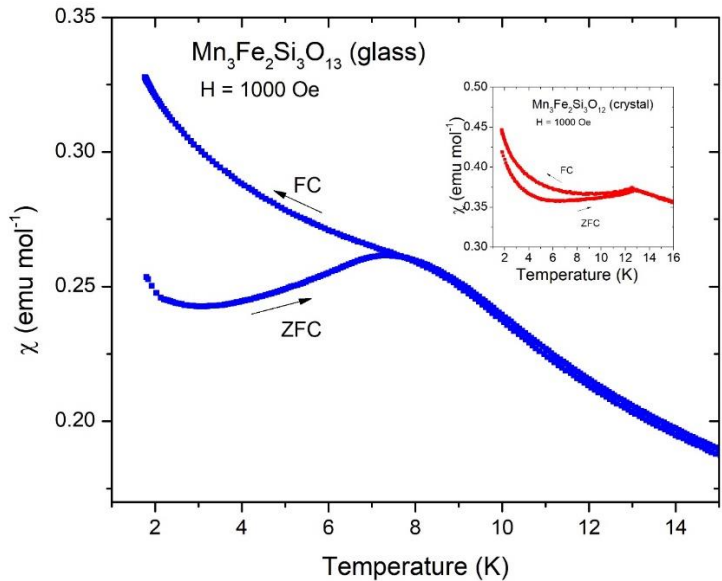

**Figure 6.** Zero-field-cooled and field-cooled DC magnetic susceptibility of the glassy phase $Mn_3Fe_2Si_3O_{13}$ in an applied field of 1,000 Oe, showing clear spin-glass behavior. Inset: Zero-field-cooled and field-cooled DC magnetic susceptibility of the crystalline garnet $Mn_3Fe_2Si_3O_{12}$ in an applied field of 1,000 Oe.

The magnetization vs applied magnetic field curve of the glass shows a much greater degree of curvature than that of the crystal, although again no saturation is observed. What appears to be a small amount of magnetic hysteresis was observed, and a full hysteresis curve of the glass can be seen in Figure 7. While this hysteresis suggests the presence of some ferromagnetic character in the spin-glass state, a Curie fit to the low-temperature region of the susceptibility curve (1.8 K to 2 K) results in a $\theta_W$ of -12.3 K, still suggesting an overall antiferromagnetically interacting system.

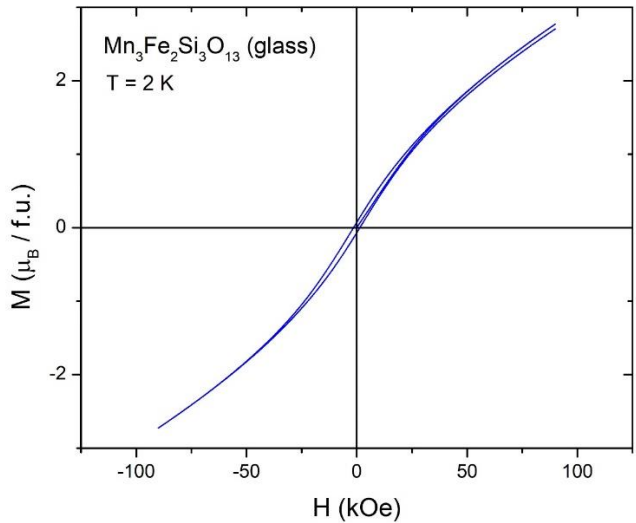

**Figure 7.** Field-dependent magnetization loop of the glassy phase $Mn_3Fe_2Si_3O_{13}$ between -90,000 and 90,000 Oe at 2 K, showing a small degree of hysteresis.

**Conclusion**

The syntheses of crystalline calderite garnet ($Mn_3Fe_2Si_3O_{12}$) and an amorphous phase of the same nominal composition have been carried out under high pressures. A profile fit of the powder XRD pattern of the crystalline phase is in good agreement with the previously published structure. The XRD pattern of the glassy phase shows only a large broad hump and a small broad peak that is believed to correspond to $SiO_2$, reflecting its amorphous nature. Thermogravimetric analysis suggests that the glassy phase has an actual composition of $Mn_3Fe_2Si_3O_{13}$, although this difference in oxidation state does not account for observed differences in the magnetic properties of the two materials. Basic magnetic measurements have been performed on both samples, revealing that they are both frustrated magnets, although the glassy phase is both a spin glass and significantly more frustrated than the crystalline phase, ordering at 2.5 K with a frustration index of approximately 33, as opposed to the crystal ordering at 12 K with a frustration index of 4. Both transitions are found to be somewhat suppressed by high fields, with the effect being more pronounced in the glassy phase. We believe that direct comparison of crystalline and glassy phases of the same nominal composition will often yield a better understanding of the magnetic interactions present, and continued work on the subject is warranted. As the toolbox of synthesis techniques continues to grow, the number of materials with both glassy and crystalline variants available for study also increases, suggesting that this will be a fruitful area of future research.

**Acknowledgments**

This research was supported by the US Department of Energy, Division of Basic Energy Sciences, Grant No. DE-FG02-08ER46544, and was performed under the auspices of the Institute for Quantum Matter.